# Transform-Limited Photon Emission From a Lead-Vacancy Center in Diamond Above 10 K

Peng Wang[1], Lev Kazak[2], Katharina Senkalla[2], Petr Siyushev[2,3,4], Ryotaro Abe[1], Takashi Taniguchi[5], Shinobu Onoda[6], Hiromitsu Kato[7], Toshiharu Makino[7], Mutsuko Hatano[1], Fedor Jelezko[2] and Takayuki Iwasaki[1, *]

[1] Department of Electrical and Electronic Engineering, School of Engineering,
Tokyo Institute of Technology, Meguro, 152-8552 Tokyo, Japan

[2] Institute for Quantum Optics, Ulm University, Albert-Einstein-Allee 11, D-89081 Ulm, Germany

[3] 3rd Institute of Physics, University of Stuttgart, Pfaffenwaldring 57, 70569 Stuttgart, Germany

[4] Institute for Materials Research (IMO), Hasselt University, Wetenschapspark 1, B-3590 Diepenbeek, Belgium

[5] Research Center for Materials Nanoarchitectonics, National Institute for Materials Science, 305-0044 Tsukuba, Japan

[6] Takasaki Advanced Radiation Research Institute, National Institutes for Quantum Science and Technology, 1233 Watanuki, Takasaki, 370-1292 Gunma, Japan

[7] Advanced Power Electronics Research Center, National Institute of Advanced Industrial Science and Technology, Tsukuba, 305-8568 Ibaraki, Japan

*Email: iwasaki.t.aj@m.titech.ac.jp

Abstract:

Transform-limited photon emission from quantum emitters is essential for high-fidelity entanglement generation. In this study, we report the coherent optical property of a single negatively-charged lead-vacancy (PbV) center in diamond. Photoluminescence excitation measurements reveal stable fluorescence with a linewidth of 39 MHz at 6 K, close to the transform-limit estimated from the lifetime measurement. We observe four orders of magnitude different linewidths of the two zero-phonon-lines, and find that that the phonon-induced relaxation in the ground state contributes to this huge difference in the linewidth. Due to the suppressed phonon absorption in the PbV center, we observe nearly transform-limited photon emission up to 16 K, demonstrating its high temperature robustness compared to other color centers in diamond.

Color centers in diamond serve as prospective quantum repeaters in large-scale quantum networks [1–4], owing to their promising performance in the coherent manipulation of spin states [5,6] and efficient quantum light-matter interfaces [7,8]. Although two-photon interference [9] and quantum entanglement [3,10,11] based on the nitrogen-vacancy (NV) center have been reported thus far, the low fraction of emission into the zero-phonon-line (ZPL) and instability due to large spectral diffusion pose a challenge for constructing a large-scale quantum network based on NV centers [12–14]. Alternatively, inversion-symmetric negatively-charged group-IV vacancy centers, including silicon-vacancy (SiV), germanium-vacancy (GeV), tin-vacancy (SnV), and lead-vacancy centers (PbV), demonstrate large fluorescence concentrations in their ZPLs and high robustness against external noise [15,16]. In particular, the PbV center is predicted to exhibit a millisecond spin coherence time at 9 K [17], making it promising for quantum applications.

The Pb atom, composing a PbV center, is the largest and heaviest among the stable group-IV elements, leading to severe lattice damage during ion implantation and difficulty in accommodating the Pb atom in the interstitial split-vacancy configuration in the diamond crystal; indeed, different ZPL wavelengths and large background emissions have been observed [18–21]. In our previous work [17], the high-temperature annealing over 2000°C under high-pressure has been applied for the formation of PbV centers, leading to the determination of the ZPL wavelengths. However, so far, the photon emission with the transform-limited linewidth remains an open challenge, which is essential for high photon interference visibility in quantum networks [22,23]. Mitigation of phonon-induced relaxation and spectral diffusion plays an important role in achieving transform-limited photons. In this work, we investigate the optical properties of single PbV centers under resonant excitation and demonstrate that the C-transition, one of the ZPLs, reaches the nearly transform-limit at 6.2 K without prominent phonon-induced relaxation and spectral diffusion. On the other hand, the D-transition, another ZPL, shows a linewidth four orders of magnitude broader than that of the C-transition. We show that this stark contrast between the two ZPLs originates from phonon-induced relaxation in the ground state of the PbV center. Furthermore, the suppressed phonon relaxation on the C-transition enables us to observe the nearly transform-limited photons up to 16 K, in contrast to other group-IV vacancy centers in diamond.

The PbV centers are fabricated in a CVD-grown IIa-type diamond substrate. Ion implantation at an acceleration energy of 330 keV leads to a projected depth of 57 nm from the surface [17] according to SRIM calculations [24]. Subsequent high-pressure and high-temperature annealing (7.7 GPa, 2100°C) leads to the formation of PbV centers and simultaneously recover the lattice damage generated during heavy-ion implantation [25,26]. The experiments are carried out on cryogenic home-built confocal microscope systems at varying temperatures. Photoluminescence (PL) measurements are carried out with a spectrometer with 532 nm laser excitation. Photoluminescence excitation (PLE) measurements are performed with a tunable dye laser and a wavemeter while measuring the phonon-side-band (PSB).

The fluorescence detection for mapping and PLE are conducted based on the Qudi python module [27].

The PbV center in diamond possesses the same crystal and electronic structure as other group-IV vacancy centers; it contains an interstitial Pb atom and two neighboring vacancies, leading to $D_{3d}$ symmetry with the main axis oriented along the ⟨111⟩ direction [28], as shown in Fig. 1(a). The large atomic size of Pb leads to poor accommodation in the diamond lattice, and thus, a rather severe strain environment is expected compared with other color centers [28]. The negatively-charged PbV has 11 electrons, leading to a spin-1/2 system with a four-level energetic structure (Fig. 1(b)). The spin-orbit interaction induces split ground ($\omega_{GS}/2\pi$) and excited ($\omega_{ES}/2\pi$) states, producing four ZPLs, labeled A, B, C, and D in the sequence of the energy decrease. The phonon absorption and emission between the sublevels in the ground state are depicted as dashed arrows, which will be discussed later. Figure 1(c) shows a PL spectrum at room temperature (~287 K) with 532 nm excitation. The prominent emissions at 552 nm and 556 nm correspond to the C- and D-transitions, respectively, followed by the PSB emission centered at 585 nm. The two ZPLs comprise ~30% of the total fluorescence, which is comparable to the Debye Waller factor of the SnV center [26]. Although the concentration to ZPL is lower compared to the SiV [29,30] and GeV [31] centers due to the lattice distortion generated with the heavy atom in diamond [28], it is still ten times larger than that of the NV center in diamond [32]. The inset in Fig. 1(c) shows the fine structure of the emission, recorded at 6.2 K; the linewidth of the C-transition is limited to the spectrometer resolution of ~50 GHz, while the D-transition exhibits a much broader linewidth over 400 GHz. The ground state splitting is 3870 GHz estimated from the energy difference of the two peaks. The wavelengths of the A- and B-transitions predicted according to the excited state splitting (6920 GHz) from *ab initio* calculations [28] are indicated with blue arrows. They are not observed here due to insufficient thermal excitation into the upper branch in the excited state [17].

Then, we explore the coherent optical properties of a single PbV center in diamond. A time-resolved photoluminescence (TRPL) measurement is performed with a 532 nm pulsed laser. From the single exponential decay curve in Fig. 2(a), the radiative lifetime of the PbV center is estimated to be 4.4±0.1 ns, leading to a transform-limited linewidth of 36.2±0.8 MHz. A 550 nm bandpass filter used for TRPL enables the detection of both C- and D-transitions, and thereby, we assume that both have the same transform-limit, as similar to SiV centers [33,34]. The PLE measurement is carried out by tuning the excitation laser, while detecting the PSB with a 561 nm longpass filter. Figure 2(b) shows a PLE spectrum of the C-transition from the identical PbV center for the lifetime measurement above. The single-scan PLE spectrum under 1 nW resonant excitation demonstrates a narrow Lorentzian peak with a linewidth of 38.8±0.3 MHz, which agrees well with the transform-limited linewidth. Thus, the radiative transition plays a dominant role in the linewidth of the C-transition of the PbV center at 6.2 K. During the repeated scans on the target single PbV center, no prominent peak shift is observed over time, as shown in Fig. 2(c). Averaging all PLE spectra yields a linewidth of 40.6 MHz, indicating that spectral diffusion is greatly suppressed in the PbV center formed by high-temperature annealing. Notably, we occasionally observe a sudden termination of the emission close to zero detuning (Supplementary Materials). It is related to the charge conversion and has been reported in other group-IV vacancy centers [35–37]. A non-resonant 532 nm laser or even a resonant laser itself can serve as a repump laser to recover the emission (Supplementary Materials). Control of the charge state of the PbV center is of vital importance for future quantum applications.

The TRPL and PLE measurements are conducted on multiple PbV centers. Note that these two types of measurements are not performed for the same PbV centers. The distributions of the radiative lifetime and linewidth of the C-transition are shown in Fig. 2(d). The mean lifetime of 4.4 ns corresponds to a transform-limited linewidth of 36 MHz, and matches the mean linewidth of 39 MHz obtained from the single-scan PLE measurements. Furthermore, the radiative lifetime of a PbV center at varying temperatures is measured. The lifetime is nearly constant from a cryogenic temperature to 250 K, as shown in Fig. 2(e). This fact suggests a relatively high activation energy for non-radiative decay similar with the GeV center [38], indicating a high quantum efficiency of the PbV center in diamond.

In contrast to the coherent C-transition, another optical transition, the D-transition, shows significant broadening over 400 GHz even at low temperature (inset in Fig. 1(a)). This result is completely different from the observation on the SiV center [33], where the two transitions exhibit similar narrow linewidths within 1.5 times the transform-limit. For further comparison to another type of color center, we measure the PLE spectra of a GeV center (Fig. 3(a)). The linewidths are 53.6 and 98.3 MHz for the C- and D-transitions, respectively, at 5 nW resonant excitation. A transform-limited linewidth of 28.9 MHz is obtained from the radiative lifetime (Supplementary Materials), and a power broadening of 9.6 MHz is estimated from the C-transition at 1 nW excitation (Fig. 3(a)). Again, the slight broadening of the D-transition on the GeV center does not match with the behavior of the PbV center.

To understand the measurement results, the relaxation processes in the group-IV vacancy centers are considered. The line broadening on a quantum emitter can be formulated as follows [22,39–41]:

$$\Gamma = \Gamma_0 + \Gamma_{ph} + \Gamma_{others} \tag{1}$$

where $\Gamma_0$ refers to the transform-limited linewidth directly obtained from the radiative lifetime, $\Gamma_{ph}$ is the phonon-induced relaxation, and $\Gamma_{others}$ comprises other effects such as power broadening and spectral diffusion. The power broadening is considered to be weak under low resonant power excitation. Additionally, a low non-resonant excitation power of 0.2 mW is used for PL of the D-transition in Fig. 1(c), which is 1/9 of the saturation power [17]. Spectral diffusion does not play a significant role, as shown in Fig. 2(c). Conclusively, the broadening of the D-transition in the PbV center is mainly attributed to the phonon-induced relaxation between the split energy levels.

The phonon-induced transition occurs in both the excited and ground states, where the latter is depicted in Fig. 1(b). In the ground state, phonon absorption and emission contribute to the line broadening of the C- and D-transitions, respectively. Due to the E-symmetry electronic structure in the group-IV vacancy center in diamond, only the electron-phonon coupling with the E-symmetric acoustic phonon modes can be considered [42,43]. We assume that single phonon processes, resonant to the split energies in the ground and excited states, are dominant at low temperatures. According to the Debye model and Fermi's golden rule, the phonon-induced transition rate is given as follows [39,43,44](also see Supplementary Materials):

$$\begin{cases} \gamma_s^+ = 2\pi\alpha\omega_s^3 n(\omega_s, T) \\ \gamma_s^- = 2\pi\alpha\omega_s^3 [n(\omega_s, T) + 1] \end{cases} \quad (s = GS, ES) \quad (2)$$

where $\gamma_s^+$ and $\gamma_s^-$ represent the phonon absorption and emission rates, respectively, for a resonant phonon frequency of $\omega_s$. GS and ES represent the ground state and excited state, respectively. $n(\omega_s, T)$ is the mean number of phonons at thermal equilibrium and defined as follows: $n(\omega_s, T) = [\exp(\hbar\omega_s/(k_B T)) - 1]^{-1}$, where $k_B$ is the Boltzmann constant. $\alpha$ is a coupling parameter. Clearly, the phonon emission rate ($\gamma_s^-$) is larger than the phonon absorption rate ($\gamma_s^+$) for a given phonon frequency and temperature. Considering the electron-phonon interactions in both ground and excited states, Equations (1) and (2) lead to the linewidths of the C- and D-transitions as follows:

$$\begin{cases} \Gamma_C = \Gamma_0 + \Gamma_{others} + \frac{1}{2\pi}(\gamma_{GS}^+ + \gamma_{ES}^+) \\ \Gamma_D = \Gamma_0 + \Gamma_{others} + \frac{1}{2\pi}(\gamma_{GS}^- + \gamma_{ES}^+) \end{cases} \quad (3)$$

Importantly, the C-transition is affected only by the phonon absorption in both the ground and excited states ($\gamma_{GS}^+$ and $\gamma_{ES}^+$), while the D-transition contains the term on the phonon emission in the ground state ($\gamma_{GS}^-$). We obtain a relationship of the linewidth difference between the C- and D-transitions from Equations (2) and (3):

$$\Gamma_D - \Gamma_C = \frac{1}{2\pi}(\gamma_{GS}^- - \gamma_{GS}^+) = \alpha\omega_{GS}^3 \quad (4)$$

This equation clearly indicates that the D-transition is broader than the C-transition independent of temperature. The difference in the linewidth depends on the ground state splitting, which increases as increasing the size of the group-IV element due to the stronger spin-orbit interaction [16].

The experimental linewidth differences of the GeV and PbV centers are well fitted with this cubic trend on the ground state splitting (Fig. 3(b)), with the coefficient $\alpha$ of $(2\pi)^{-3} \times 7.51 \times 10^{-9}$ GHz$^{-2}$. Therefore, the phonon-induced relaxation model above can explain the large broadening of the D-transition in the PbV center. Note that the effect of the relaxation in the excited state ($\gamma_{ES}^+$) is rather small due to the large spin-orbit interaction (Supplementary Materials). Notably, according to the phonon relaxation model in this study, the SiV center has a very small linewidth difference below 1 MHz (Fig. 3(b)), which is challenging to experimentally observe. Moreover, the D-transition of SnV is not observed in our PLE measurements, probably due to the low intensity of the predicted broad linewidth of approximately 4 GHz.

Finally, we discuss the temperature dependence of the coherent optical properties of the group-IV vacancy centers in diamond based on phonon-induced relaxation. Figure 4(a) shows the PLE spectra of the C-transition of the PbV center at different temperatures. The linewidths are summarized in Fig. 4(b), together with the data of the GeV center from Fig. 3(a) and a SnV center (Supplementary Materials) for comparison. Note that the linewidths are normalized to the transform-limited linewidth for each color center: 28.9 MHz for GeV (Supplementary Materials), 30.6 MHz for SnV [45], and 36.2 MHz for the PbV center. Surprisingly, the nearly transform-limited photon from the PbV center is observed above 10 K, indicating the weak phonon-induced relaxation of the C-transition even at elevated temperatures. The solid lines represent the linewidths of the C-transition of the SnV and PbV centers calculated by Equation (3) with a fitting parameter $\Gamma_{others}$, assuming a negligible contribution of phonon relaxation in the excited state (Supplementary Materials). The calculated values agree with the experimental observations below 20 K. The discrepancy at higher temperatures could be attributed to higher-order electron-phonon interactions [43,46].

Line broadening beyond the transform-limit causes indistinguishability degradation in two-photon interference. With an emission linewidth of approximately 1.2 times the transform-limit, a visibility of ~80% is obtained for two-photon interference [22]. To satisfy this condition, according to the phonon-induced relaxation model, the SiV and GeV centers require cooling to 4~5 K, while the temperature tolerance of the SnV center is 6 K. In contrast, the estimated temperature for the PbV center reaches 16 K, which agrees with the experimental observation (Fig. 4b). This temperature is also higher than an estimated temperature of ~11 K for the NV center in diamond [47,48]. This superior thermal characteristic opens great opportunity to build energy-efficient scalable quantum networks based on the PbV center in diamond.

In conclusion, we obtained transform-limited photon emission without evident spectral diffusion from the C-transition of a single PbV center in diamond. The coherent optical transitions for multiple PbV centers were confirmed. The large difference in the linewidth on the C- and D- transitions are well explained by considering the phonon-induced relaxation. Owing to its large ground state splitting, the PbV center displayed high temperature robustness of the coherent optical property up to 16 K, much higher than those for other group-IV vacancy centers in diamond. These results pave the way for the PbV center to become building blocks to construct large-scale quantum networks.

The authors would like to thank Masahiko Ogura, Tuan Minh Hoang, and Kohei Suda for supporting in the sample preparation, and Yasuyuki Narita, Kazuki Oba, and Yoshiyuki Miyamoto for fruitful discussions. This work is supported by JSPS KAKENHI (Grants No. JP22H04962, No. JP22H00210, and No. JP23KJ0931), the Toray Science Foundation, the MEXT Quantum Leap Flagship Program (Grant No. JPMXS0118067395), and JST Moonshot R&D (Grant No. JPMJMS2062). Fedor Jelezko acknowledges support of DFG via projects 386028944, 445243414, 387073854, BMBF via project SPINNING, CoGeQ and QRX, ERC via Synergy Grant HyperQ.

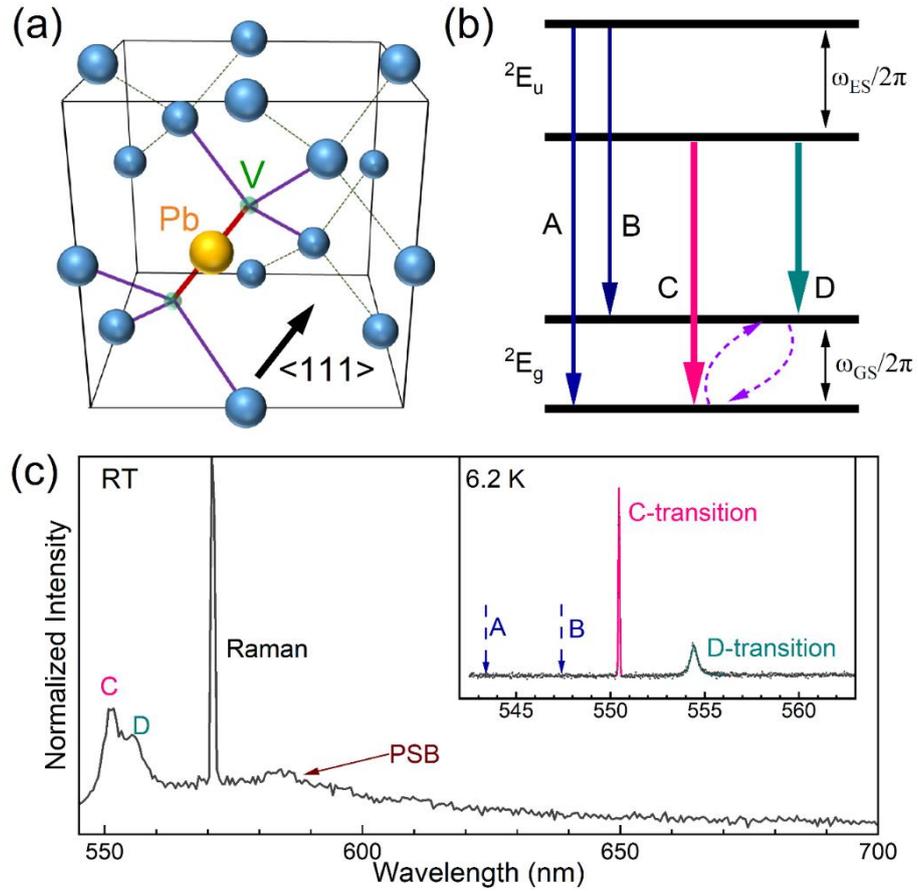

Fig. 1 Atomic structure and optical characteristics of the PbV center in diamond. (a) Crystal structure of the PbV center. (b) Four-level energetic structure of the PbV center and the corresponding optical transitions, labeled A, B, C and D. The phonon-induced relaxation processes (phonon absorption and emission) in the ground states are shown by dashed arrows. (c) PL spectra of the PbV center at room temperature (RT) and 6.2 K (inset). The peaks at 550 nm (pink) and 554 nm (cyan) are the C- and D-transitions, respectively. The predicted A- and B-transitions are marked with blue arrows.

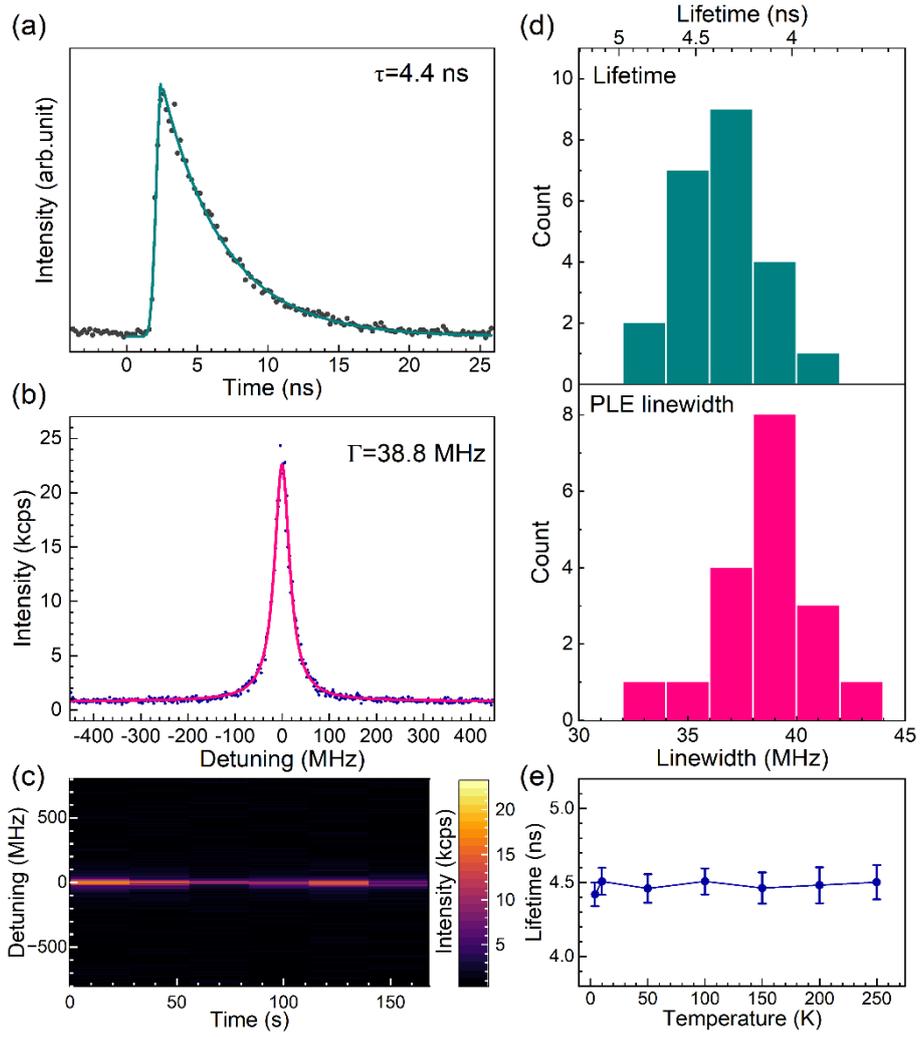

Fig. 2 Transform-limited emissions from PbV centers. (a) Time-resolved photon emission. (b) PLE spectrum of the C-transition at 1 nW resonant excitation. (c) PLE spectra as a function of time. (d) Distribution of the radiative lifetime and single-scan PLE linewidth of multiple PbV centers. The top axis of the lifetime distribution is converted to the transform-limited linewidth (bottom axis). (e) Excited-state lifetime as a function of temperature.

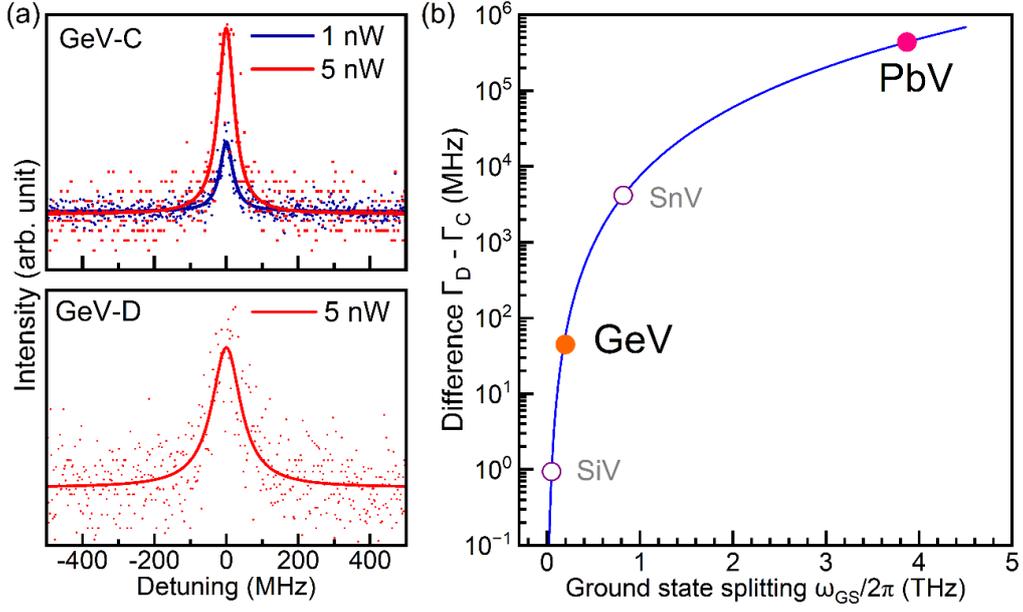

Fig. 3 Linewidth difference of the C- and D-transitions of group-IV vacancy centers in diamond. (a) PLE spectra of the C- and D-transitions of a GeV center at 6.2 K. (b) Linewidth difference between C- and D-transitions ($\Gamma_D - \Gamma_C$) as a function of the ground state splitting. The solid circle represents the experimental results of the GeV and PbV centers. The fitting is performed based on the phonon-induced relaxation model in Equation (4) for GeV and PbV. The hollow circles are plotted according to the ground state splitting of the SiV and SnV centers.

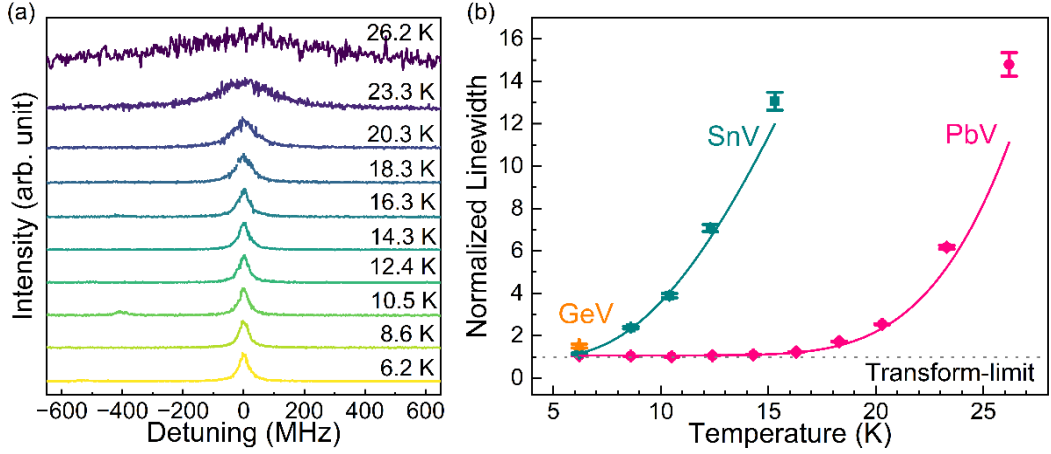

Fig. 4 Temperature dependence of the C-transition. (a) PLE spectra of the C-transition of the PbV center recorded at temperatures from 6.2 to 26.2 K. All spectra are obtained from three repeated scans. The peak centers are aligned for clarity. (b) Temperature-dependent linewidth of the C-transition of group-IV vacancy centers in diamond. The linewidths are normalized to each transform-limit (28.9 MHz for GeV, 30.6 MHz for SnV [45], and 36.2 MHz for PbV). The solid lines represent the calculations based on the phonon-induced relaxation model with the parameter $\alpha$ obtained in Fig. 3(b) and fitting parameters of $\Gamma_{\text{others}}(\text{PbV}) = 2.7\ \text{MHz}$ and $\Gamma_{\text{others}}(\text{SnV}) = -1.8\ \text{MHz}$. The negative value for SnV is attributed to a possibly longer radiative lifetime of the target SnV compared to that of the ensemble state [45].

# Supplementary Materials
# for Transform-Limited Photon Emission From a Lead-Vacancy Center in Diamond Above 10 K

Peng Wang[1], Lev Kazak[2], Katharina Senkalla[2], Petr Siyushev[2,3,4], Ryotaro Abe[1], Takashi Taniguchi[5], Shinobu Onoda[6], Hiromitsu Kato[7], Toshiharu Makino[7], Mutsuko Hatano[1], Fedor Jelezko[2] and Takayuki Iwasaki[1]

[1] Department of Electrical and Electronic Engineering, School of Engineering, Tokyo Institute of Technology, Meguro, 152-8552 Tokyo, Japan

[2] Institute for Quantum Optics, Ulm University, Albert-Einstein-Allee 11, D-89081 Ulm, Germany

[3] 3rd Institute of Physics, University of Stuttgart, Pfaffenwaldring 57, 70569 Stuttgart, Germany

[4] Institute for Materials Research (IMO), Hasselt University, Wetenschapspark 1, B-3590 Diepenbeek, Belgium

[5] Research Center for Materials Nanoarchitectonics, National Institute for Materials Science, 305-0044 Tsukuba, Japan

[6] Takasaki Advanced Radiation Research Institute, National Institutes for Quantum Science and Technology, 1233 Watanuki, Takasaki, 370-1292 Gunma, Japan

[7] Advanced Power Electronics Research Center, National Institute of Advanced Industrial Science and Technology, Tsukuba, 305-8568 Ibaraki, Japan

**I. Observation of a single PbV center in diamond**

Figure S1(a) shows confocal fluorescence mapping with a 550 nm bandpass filter to detect the PbV centers. Several isolated fluorescent spots are clearly observed, where the target single PbV is highlighted in a white circle. The Hanbury-Brown and Twiss (HBT) measurement [1] with a 50:50 beam splitter is performed to confirm the addressing of a single PbV center (Fig. S1(b)). Even without background correction, the antibunching at zero-time delay drops to 0.08 at 0.25 mW non-resonant excitation, indicating single photon emission [2] from the target PbV center.

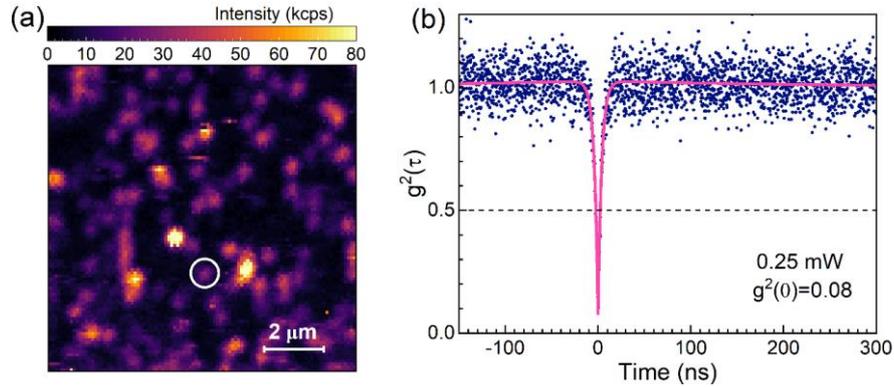

Fig. S1 Observation of a single PbV center in diamond. (a) Confocal fluorescence mapping with a 550 nm bandpass filter, where the target PbV center is marked in a white circle. (b) Second-order autocorrelation of the PbV center.

**II. Charge state conversion**

In previous works, the other group-IV vacancy centers end up in a non-radiative state under resonant excitation, which is attributed to the possible charge conversion of the SiV, GeV, and SnV centers in diamond [3–5]. Similar emission extinction is also observed in the PbV center, as shown in Fig. S2(a). During the repeated PLE measurements, the intensity suddenly drops when approaching zero detuning. The charge state can be restored with a non-resonant excitation; here, we introduce a laser power of 0.2 mW at 532 nm. After the repump process, a clear PLE peak appears again with almost no spectral diffusion.

Interestingly, we observe that the resonant laser itself (~550 nm) can serve as a repump laser, resulting in the cycling of extinction and recovery during the PLE measurements on another PbV center (Fig. S2(b)). After beaching of the center while scanning through the resonance, the luminescence suddenly recovers (top panel). The spectral diffusion is observed during the consecutive scans under resonant excitation of this emitter (bottom panel). This observation indicates that a higher number of defects, such as divacancy [5], which provides charges for recovery, exist around this emitter and that even weak resonant excitation functions as a repump laser to excite the defects, leading to a significant change in the charge environment and subsequent spectral diffusion. The stability and charge recovery are believed to be highly dependent on the surrounding environment of the specific PbV center, causing emitter-to-emitter behaviors.

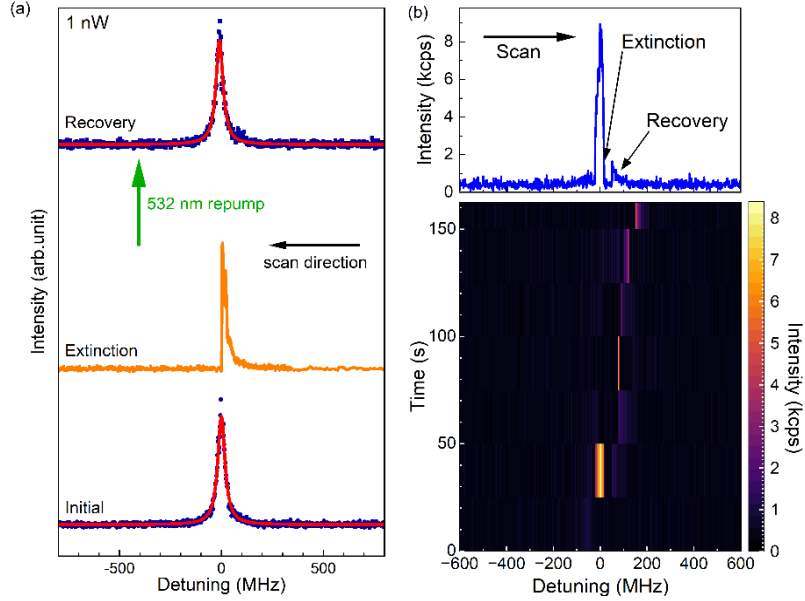

Fig. S2 Charge conversion of PbV centers. (a) Fluorescence extinction and recovery of the PbV center. The initial PLE spectrum is shown at the bottom. The repeated measurements on this PbV center lead to the extinction of emission (middle). The emission recovery is observed after 532 nm laser repump (top). (b) The resonant laser-induced repumping. The top panel is the second scan of the bottom panel.

**III. Fabrication of the GeV and SnV centers**

The SnV centers are fabricated in a IIa-type (001) single-crystal diamond substrate [6]. Sn ion is implanted with an acceleration energy of 18 MeV with a fluence of $5 \times 10^8 \text{cm}^{-2}$. The projected depth is approximately 3 μm. The annealing is carried out at 2100°C under a high pressure of 7.7 GPa for 20 min. The GeV centers are fabricated in a lightly phosphorus-doped ([P]~ $10^{16}\text{cm}^{-3}$) diamond layer grown on a (111) single-crystal diamond substrate. Ge ion implantation is performed at an acceleration energy of 700 keV with a fluence of $1 \times 10^{10}$ cm$^{-2}$, producing a projected depth of approximately 270 nm. The sample is annealed in vacuum at 1000°C for 60 min after ion implantation.

**IV. Optical properties of the GeV and SnV centers**

The GeV and SnV centers show ZPL at 602 nm and 619 nm, respectively [7,8]. The TRPL measurement is conducted on the GeV center (Fig. S3). Different from the PbV center in the main text, the intensity curve is fitted with a biexponential decay to consider the sharp decay from 1 to 2 ns, which comes from background emission, such as the second-order Raman of diamond [6]. The radiative lifetime is estimated to be 5.5 ns, giving rise to a transform-limited linewidth of 28.9 MHz, which agrees with previous reports [9,10].

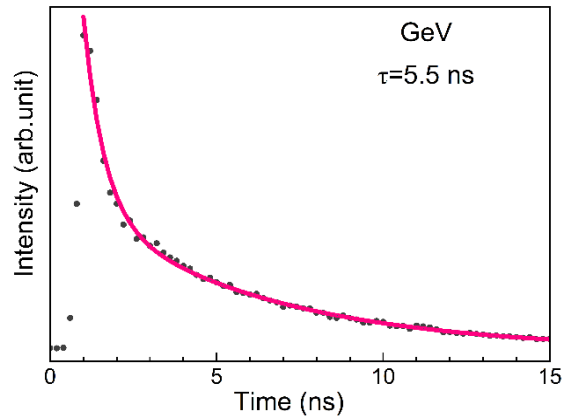

Fig. S3 Lifetime measurement of the GeV center in diamond.

Figure S4 shows the PLE spectra of the C-transition of the SnV center from 6.2 to 15.3 K. A narrow linewidth of 35.4 MHz is obtained at 6.2 K. Moreover, an increase in temperature above 8 K leads to significant broadening, which is in contrast to the observation on the PbV center.

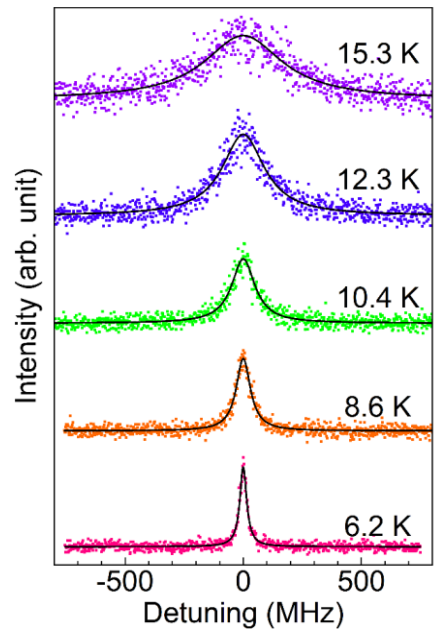

Fig. S4 PLE spectra of the SnV center at various temperatures.

## V. Phonon-induced relaxation

Phonon-induced relaxation is obtained based on the electron-phonon interaction [11,12]. The Hamiltonian of the acoustic phonon can be written with creation and annihilation operators.

$$\hat{H}_{ph} = \sum_k \hbar\omega_k (\hat{a}_k^\dagger \hat{a}_k) \tag{S1}$$

where $\omega_k$ is the angular frequency of the acoustic phonon. The electron-phonon coupling is described as follows:

$$\hat{H}_{e-ph} = \sum_k \lambda_{ij,k} |i\rangle\langle j| (\hat{a}_k^\dagger + \hat{a}_k) \tag{S2}$$

where $\lambda_{ij,k}$ represents the linear coupling constant between electronic states of $|i\rangle$ and $|j\rangle$ for the phonon with wavevector k. Here, we consider the single phonon relaxation under the resonant condition, i.e., $\omega_k = \omega_{ij} = |E_i - E_j|/\hbar$. The linear coupling constant $\lambda_{ij,k}$ can be described as $\lambda_{ij,k}^2 \propto \omega_{ij}/\omega_D$ for $0 \ll \omega_{ij} \ll \omega_D$ [12–15], where $\omega_D$ is the Debye frequency of diamond. According to the Debye model, the phonon density is given as $\rho(\omega_{ij}) \propto \omega_{ij}^2$. Then, based on Fermi's golden rule, we obtain the single phonon-induced relaxation rate in the two-level system with a coefficient $\alpha$.

$$\begin{cases} \gamma_{ij}^+ = 2\pi\alpha\omega_{ij}^3 n(\omega_{ij}, T) \\ \gamma_{ij}^- = 2\pi\alpha\omega_{ij}^3 [n(\omega_{ij}, T) + 1] \end{cases} \tag{S3}$$

In the group-IV vacancy centers in diamond, the electronic states of $|i\rangle$ and $|j\rangle$ correspond to the split ground and excited states, forming Equation (2) in the main text.

The split frequencies in the ground and excited states as well as the transform-limited linewidths of the group-IV vacancy centers in diamond are summarized in Table S1.

Table S1. Ground state (GS) and excited state (ES) splitting and transform-limited linewidth of the group-IV vacancy centers in diamond.

|  | SiV | GeV | SnV | PbV |
| --- | --- | --- | --- | --- |
| GS splitting $\omega_{GS}/2\pi$ (GHz) | 50 [16] | 200 | 821 [6] | 3870 |
| ES splitting $\omega_{ES}/2\pi$ (GHz) | 260 [16] | 1120 [17] | 3000 [8] | 6920 [18] |
| Transform-limit (MHz) | 92.5 [19] | 28.9 | 30.6 [6] | 36.2 |

We plot the phonon-induced broadening of the linewidth with the fitting parameter $\alpha$ from the main text (Fig. S5). For the lightest SiV center in diamond, $\Gamma_{GS}^-$ and $\Gamma_{GS}^+$ have similar values at temperatures over 4 K, leading to comparable linewidths of the C- and D-transitions; in addition, the phonon relaxation in the ground state is so weak that the temperature-dependent broadening is dominated by phonon absorption in the excited state, $\Gamma_{ES}^+$, which is consistent with the discussion in a previous study [11]. However, the D-peak broadening in Ref. [11] have deviations from our calculation; these deviations are caused by the different coupling parameter $\alpha$ in the excited state from that in the ground state owing to the different coupling constants of the electron-phonon interaction. By comparing the fit result in Ref. [11] with Equation (3) in the main text, a coupling parameter $\alpha$ in the excited state is derived as $(2\pi)^{-3} \times 1.75 \times 10^{-8}$ GHz$^{-2}$, which is almost double the parameter in the ground state. In contrast to the SiV center, $\Gamma_{ES}^+$ is negligible in the SnV and PbV centers owing to the large excited state splitting (Fig. S5). Therefore, the broadening of the C- and D-transitions of those color centers can only be determined by $\Gamma_{GS}^+$ and $\Gamma_{GS}^-$, respectively.

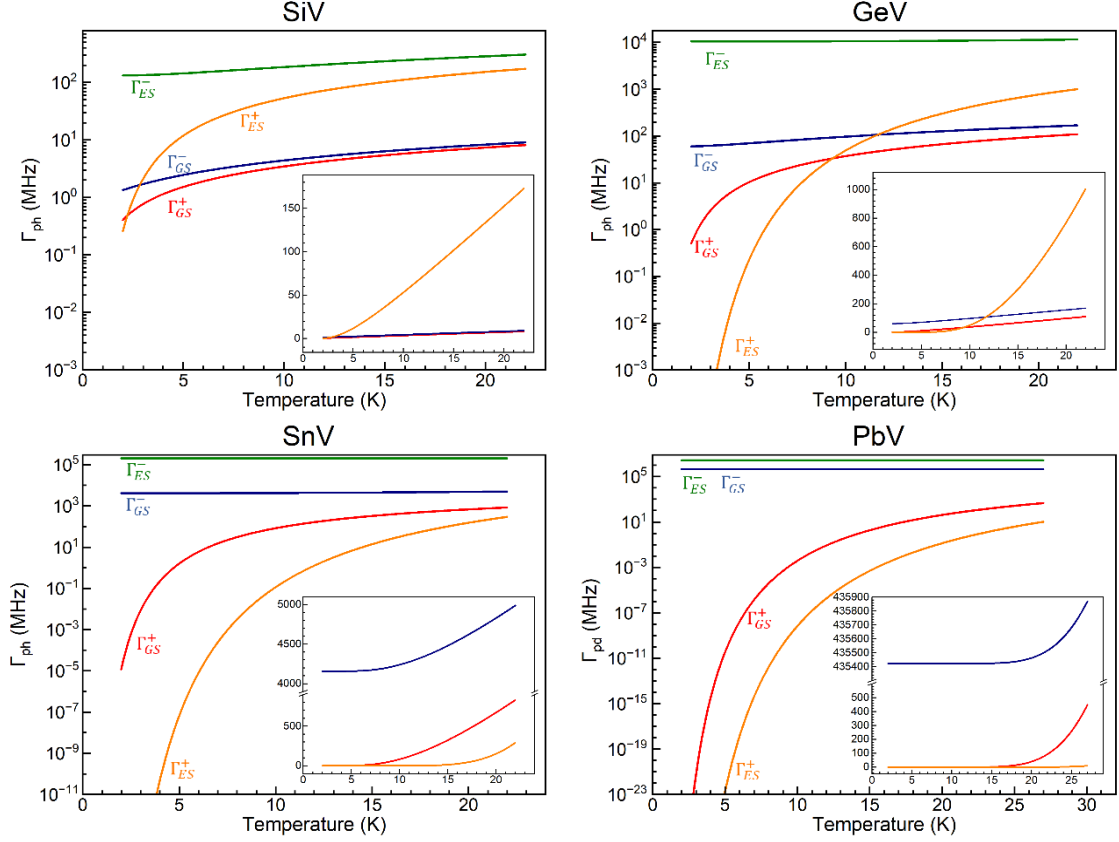

Fig. S5 Line broadening due to phonon-induced relaxation in the ground and excited states of group-IV color centers in diamond. $\Gamma^+$ and $\Gamma^-$ represent the linewidth broadening due to phonon absorption and emission, respectively. The calculations are performed with the coefficient $\alpha$ of $(2\pi)^{-3} \times 7.51 \times 10^{-9}$ GHz$^{-2}$, obtained in the main text.